\documentclass[a4paper,11pt]{article}
\usepackage{pos}

\newcommand{\splus}{\ensuremath _{\scriptscriptstyle +}}
\newcommand{\sminus}{\ensuremath _{\scriptscriptstyle -}}

\title{Topology and the Dirac Spectrum in Hot QCD}

\author*[a,b]{Tamas G.\ Kovacs}

\affiliation[a]{Department of Physics, Eotvos Lorand University \\
  Pazmany Peter setany 1/A, 1117 Budapest, Hungary }

\affiliation[b]{Institute for Nuclear Research \\
 Bem ter 18/C, 4026 Debrecen, Hungary}

\emailAdd{tamas.gyorgy.kovacs@ttk.elte.hu}

\abstract{It is known that contrary to expectations, the order parameter of
  chiral symmetry breaking, the Dirac spectral density at zero virtuality
  does not vanish above the critical temperature of QCD. Instead, the spectral
  density develops a pronounced peak at zero. We show that the spectral
  density in the peak has large violations of the expected volume
  scaling. This anomalous scaling and the statistics of these eigenmodes is
  consistent with them being produced by mixing instanton and antiinstanton
  zero modes. Consequently, we show that a nonvanishing topological
  susceptibility implies a finite density of eigenvalues around zero, which
  can have implications on the restoration of chiral symmetry above the
  critical temperature.  }

\FullConference{%
  The 39th International Symposium on Lattice Field Theory (Lattice2022),\\
  8-13 August, 2022 \\
  Bonn, Germany 
}

\begin{document}
\maketitle

\section{Introduction}

The spectrum of the quark Dirac operator has long been known to provide
information about the physics of QCD. In particular, the long-distance physics
is encoded in the low-lying, infrared part of the Dirac spectrum. The spectral
quantity that has probably the simplest direct connection to the physics is
the spectral density at zero virtuality, $\rho(0)$, which -- by the
Banks-Casher relation -- is the order parameter of chiral symmetry breaking
\cite{Banks:1979yr}. Indeed, according to the standard picture of the finite
temperature transition in QCD, in the low temperature, hadronic phase
$\rho(0)$ is nonzero, signaling the spontaneous breaking of chiral
symmetry. As the system is heated and enters the high temperature, quark-gluon
plasma phase and chiral symmetry gets restored, $\rho(0)$ becomes zero.

Strictly speaking, the above picture applies only to an imaginary world, where
quarks are massless and the chiral symmetry of the Lagrangian is exact. In the
real world the quark masses are very small compared to the QCD scale, and
chiral symmetry is only approximate. As a result, the finite temperature
transition is only a crossover, and instead of jumping to zero at a well
defined critical temperature, $\rho(0)$ only rapidly decreases through the
crossover region, as the system enters the quark-gluon plasma state.

This simple picture, however, was later challenged. The first sign of a more
complicated behavior started to emerge, when in the very early days of the
overlap Dirac operator, it was noticed that in the high temperature phase of
the quenched theory the spectral density of the overlap Dirac operator
develops an unexpected sharp peak at zero virtuality
\cite{Edwards:1999zm}. Even though, this was in the quenched theory, where
chiral symmetry is not straightforward to interpret, this behavior of the
spectral density was rather counterintuitive. These simulations were performed
on rather coarse lattices ($N_t=4$, just above $T_c$), nevertheless the
tentative interpretation of the unexpected spectral peak was that it was due
to topology-related would-be zero modes of the Dirac operator, and resolving
them was made possible only by the overlap operator, thanks to its exact
chiral symmetry. We emphasize that these are not the chiral zero modes,
corresponding to the net topological charge -- those can be easily
distinguished in the overlap spectrum. Rather, the peak in the spectral
density was suggested to be due to mixing zero modes of instantons and
antiinstantons\footnote{In the present work, for simplicity we will refer to
  lumps of unit topological charge as (anti)instantons, even though close to
  $T_c$ in the high temperature phase their charge profile might not even be
  close to that of calorons. For our discussion, the only important property
  of these objects is that they are isolated lumps of topological charge of
  unit magnitude.}, eigenmodes that come in complex conjugate pairs and can be
close to zero if their splitting due to mixing is small
\cite{Schafer:1996wv}. One could also argue that these unusual small modes
could be just cutoff effects caused by the coarse lattice, or a quenched
artifact, since fluctuations of the topological charge are expected to be
suppressed by dynamical quarks. This is, however, not the case, as was shown
in several further works using much finer lattices, as well as light dynamical
quarks \cite{Alexandru:2015fxa,Alexandru:2019gdm,Kaczmarek:2021ser}. It was
even suggested that the approximate $1/\lambda$ shape of the spectral peak
signals a previously unnoticed ``phase'' of QCD, intermediate between the
hadronic and the quark-gluon plasma phase. In the present work we will show
further evidence that the initial interpretation of the spectral peak as
mixing topological near zero modes was correct.

\section{The spectral peak}

\begin{figure}
\begin{center}
\includegraphics[width=0.8\textwidth,keepaspectratio]{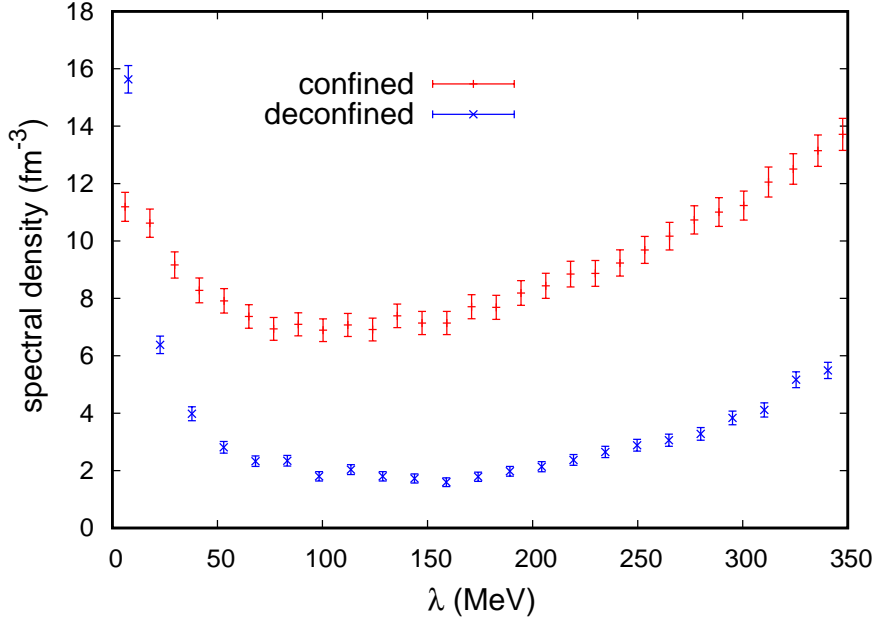}
\caption{\label{fig:spd_b6.063} The spectral density of the overlap Dirac
  operator on an ensemble of $N_t=8$ quenched lattices at the critical
  point. Based on the magnitude of the Polyakov loop, the confined and
  deconfined phases were separated, and their spectral densities are plotted
  separately.  }
\end{center}
\end{figure}

We work in the quenched approximation, but eventually will argue that our
results also apply to the case with dynamical fermions, at least in a
qualitative fashion. Let us first look at how the spectral density of the
overlap Dirac operator changes across the finite temperature phase
transition. In the quenched case with $SU(3)$ gauge group, this is a genuine
first order phase transition, and tuning the coupling to the critical point
allows us to sample both phases at the same value of the coupling. In
Fig.~\ref{fig:spd_b6.063} we show the spectral density for an ensemble of
$N_t=8$ lattices at Wilson gauge coupling $\beta=6.063$, which is the critical
value for the given temporal lattice extension. Based on the magnitude of the
Polyakov loop, we separated the configurations into two sets, belonging to the
two phases, and computed the spectral density separately in the two
phases\footnote{In the deconfined phase, only configurations in the
  ``physical''(real) Polyakov loop sector were used. In the complex sectors
  the spectral density and the properties of the corresponding eigenmodes are
  very different \cite{Kovacs:2021fwq}.}. The difference is quite clear. For
most of the spectrum the spectral density is much lower in the deconfined
phase, a behavior, consistent with the naive expectation of the spectral
density jumping to a lower value when the system enters the quark-gluon plasma
phase. However, very close to zero virtuality, the deconfined phase shows a
marked peak that goes much higher than the spectral density in the confined
phase.

\begin{figure}
\begin{center}
\includegraphics[width=0.8\textwidth,keepaspectratio]{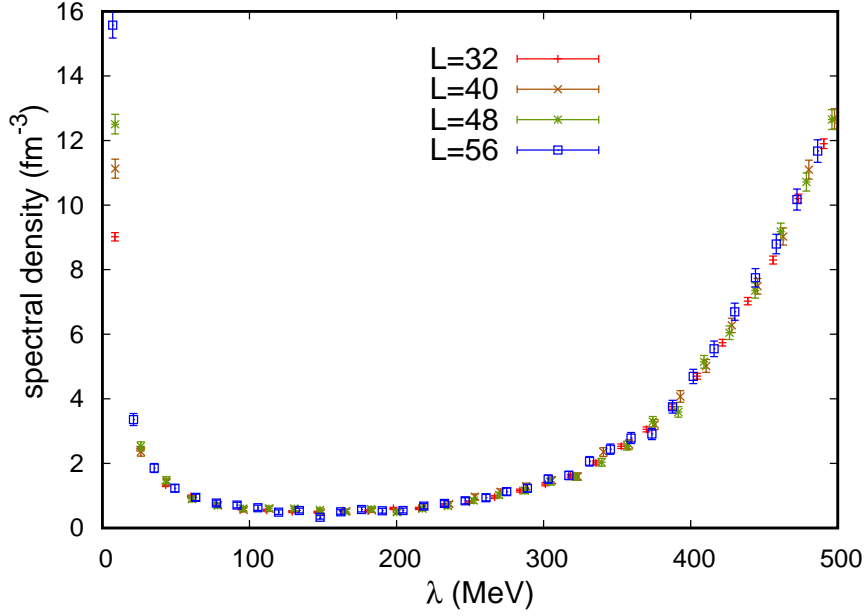}
\caption{\label{fig:spd_b6.09} The spectral density of the overlap Dirac
  operator on four ensembles of $N_t=8$ lattices at $T=1.045T_c$. The spatial
  linear lattice sizes range from $L=32$ to 56. }
\end{center}
\end{figure}

Let us further examine the peak by looking at its volume dependence. To be
able to do so in a cleaner situation where we do not have to separate the two
phases, we chose to do it at a slightly higher temperature, $T=1.045T_c$. In
Fig.~\ref{fig:spd_b6.09} we show the spectral density at $N_t=8$, Wilson
$\beta=6.09$ for ensembles of four different linear sizes, $L=32,40,48,56$. As
expected, the (volume normalized) spectral density is volume independent,
except at the spectral peak where it shows a rather strong volume dependence.

\begin{figure}
\begin{center}
  \includegraphics[width=0.49\textwidth,keepaspectratio]{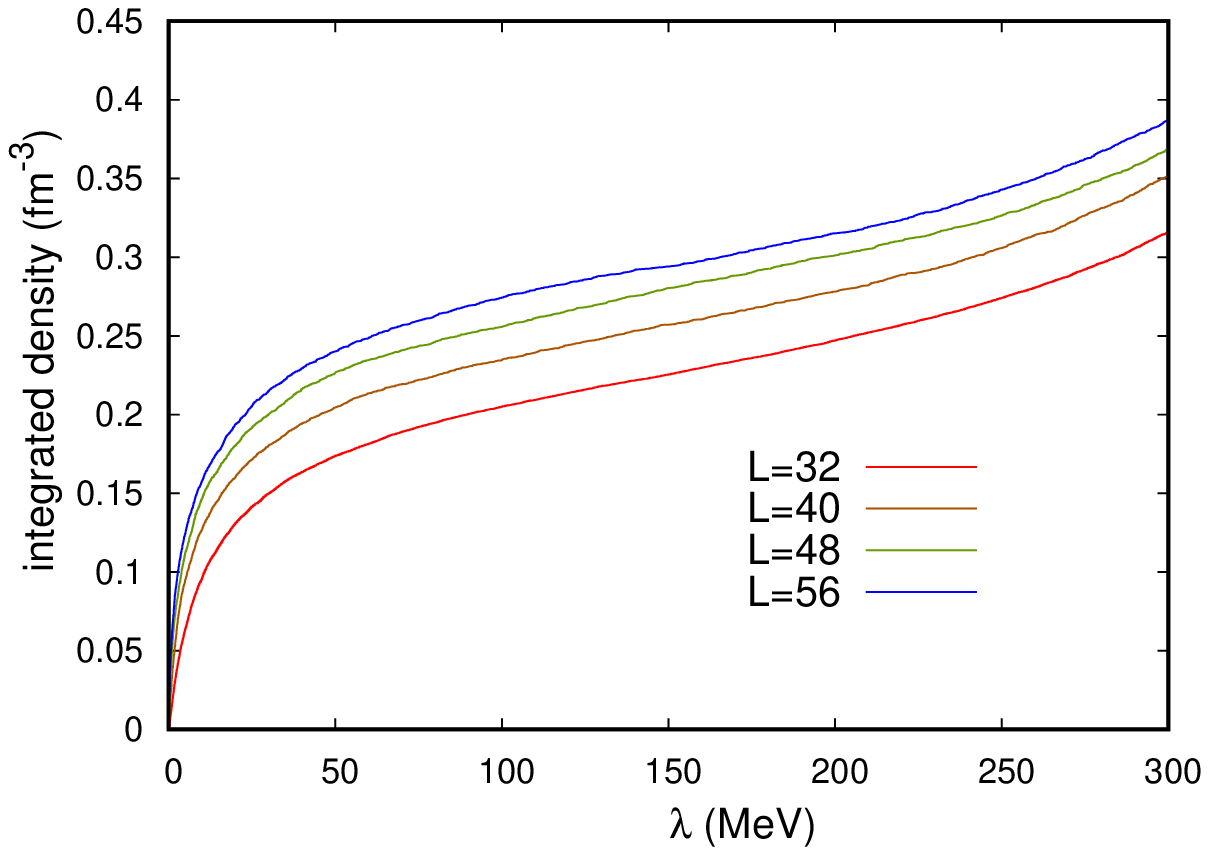}
  \hfill
  \includegraphics[width=0.49\textwidth,keepaspectratio]{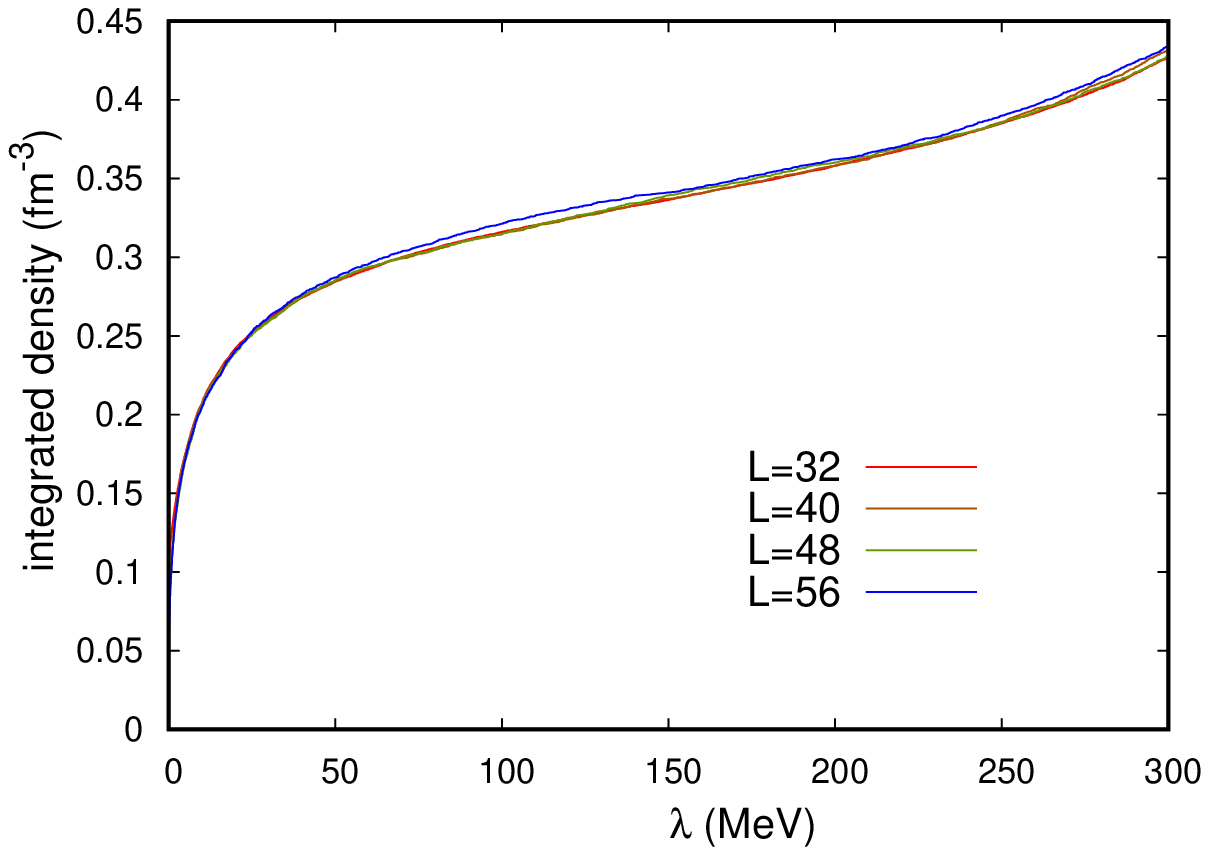}
  \caption{\label{fig:intspd} The integrated spectral density of the four
    ensembles of Fig.~\ref{fig:spd_b6.09}. In the left panel the exact zero
    modes were not included in the integral, while in the right panel they
    were. }
\end{center}
\end{figure}

The lack of volume scaling at the very low-end of the spectral density is also
manifest in the left panel of Fig.~\ref{fig:intspd}, where we show the
integrated spectral density. It is important to keep in mind that both in
Fig.~\ref{fig:spd_b6.09}, showing the spectral density and in the left panel
of Fig.~\ref{fig:intspd}, showing its integral, we did not include the exact
zero modes. In the density they would show up as a Dirac delta at the origin,
which would shift the integrated density curves upward. To check how this
affects the integrated spectral densities, in the right panel of
Fig.~\ref{fig:intspd} we also included these shifts due to the zero mode Dirac
deltas, and -- as can be seen in the figure -- the resulting curves for the
different volumes are thus shifted on top of one another. In other words, this
means that the total number of low modes scales properly with the volume, and
the ``deficit'' we observe in the near zero modes in the smaller volumes is
made up by the relatively larger density of exact zero modes there.

This indicates that the strong volume dependence of the spectral density is
connected to the zero modes and topological charge. However, these large
finite-size corrections to the density still seem quite counter-intuitive, and
it is not even clear from Fig.~\ref{fig:spd_b6.09} whether the spectral
density at zero converges to a finite value. Assuming that the modes in the
spectral peak are mixing topological zero modes, we can have a more
quantitative assessment of the expected finite-size effects in the spectral
density. Indeed, the density of topological objects is proportional to the
volume, and -- on average -- half of them are instantons, half
antiinstantons. If in a configuration there are $n\splus$ instantons and
$n\sminus$ antiinstantons, their would be zero modes combine into
$|n\splus-n\sminus|$ exact chiral zero modes, and the remaining would be zero
modes will mix into non-chiral eigenmodes close to zero, making up the
spectral peak. $\langle n\splus + n\sminus \rangle$ scales with the volume,
but
\begin{equation}
  \langle | n\splus-n\sminus | \rangle = \langle |Q| \rangle
  \propto V^{1/2},
\end{equation}
where $Q$ is the topological charge. So the volume scaling of the number of
non-chiral close to zero modes is expected to have a large finite-size
correction, vanishing in the thermodynamic limit only as $V^{-1/2}$. This is
because some of the zero modes carried by the (anti)instantons combine into
exact zero modes and do not show up in the spectral density arbitrarily close
to zero. This deficit of close to zero modes is seen in the spectral density
and causes the large finite-size correction. The missing modes, however, are
not lost, they make up the exact zero modes, and -- as we already saw -- by
their inclusion in the integrated spectral density, we can remove the large
finite-size correction.

\section{Ideal instanton gas}

Above the critical temperature, the topological susceptibility rapidly
decreases \cite{Borsanyi:2021gqg,Borsanyi:2015cka,Athenodorou:2022aay}, the
instanton gas becomes dilute and to a good approximation the (anti)instantons
form an ideal (non-interacting) gas \cite{Bonati:2013tt,Vig:2021oyt}. In such
a dilute gas, topological objects are typically far away from one another and
the splitting from the origin of the non-chiral near zero modes of topological
origin is expected to be small. Therefore, it is a reasonable assumption that
these eigenvalues of topological origin are the ones closest to zero. They
occupy a spectral region $[-\lambda_{zmz}, \lambda_{zmz}]$ that we call the
zero mode zone.

\begin{figure}
\begin{center}
\includegraphics[width=0.8\textwidth,keepaspectratio]{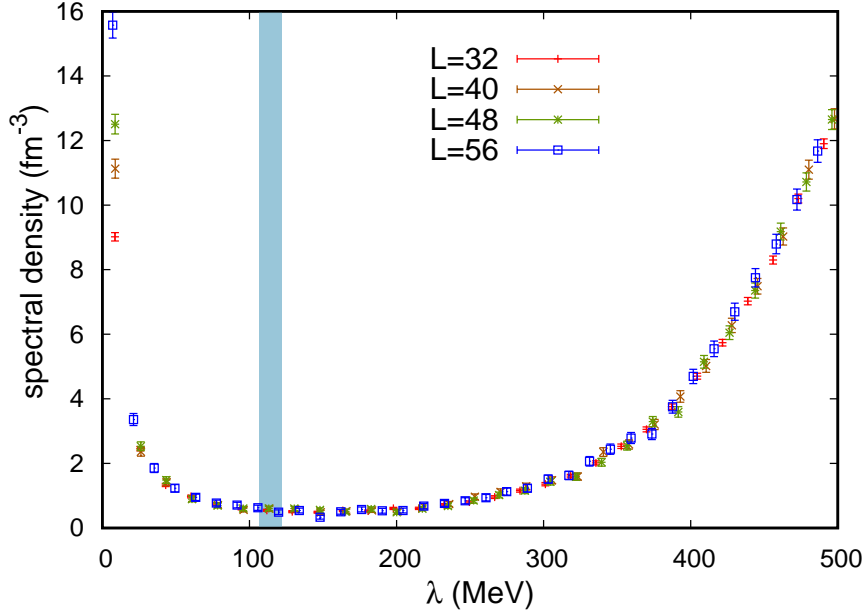}
\caption{\label{fig:spd_b6.09_zmz} The spectral density of the overlap Dirac
  operator on four ensembles of $N_t=8$ lattices at $T=1.045T_c$. The blue
  vertical bar indicates $\lambda_{zmz}$, the boundary of the zero mode zone,
  its width is the uncertainty.  }
\end{center}
\end{figure}

Using the ideal instanton gas picture, we can estimate $\lambda_{zmz}$ by
noting that in an ideal gas the number of instantons and antiinstantons follow
independent and identical Poisson distributions. The parameter of this
distribution, giving the density of topological objects is $\chi V$, where
$\chi$ is the topological susceptibility and $V$ is the space-time volume of
the system. It follows from the properties of Poisson distributions that the
density of topological objects is
\begin{equation}
  \langle n\splus + n\sminus \rangle = \chi V.
  \label{eq:poisson}
\end{equation}
From the number of exact zero modes of the overlap Dirac operator, the
topological susceptibility can be easily computed, and it is also related to
the average number of topological objects as 
\begin{equation}
  \chi V = \langle Q^2 \rangle = \langle (n\splus - n\sminus )^2 \rangle.
  \label{eq:chi}
\end{equation}
Using the susceptibility, obtained from the exact zero modes, we can estimate
$\lambda_{zmz}$ by assuming that to each instanton and antiinstanton, there is
a corresponding eigenvalue in the zero mode zone. This is done by setting
$\lambda_{zmz}$ such that on average there be $\chi V$ eigenvalues in the
$[-\lambda_{zmz}, \lambda_{zmz}]$ interval, including the exact zero modes.
In Fig.~\ref{fig:spd_b6.09_zmz} we show $\lambda_{zmz}$, the boundary of the
zero mode zone, computed in the above described way. As can be seen in the
figure, the zero mode zone includes the whole of the spectral peak and the
lower part of the wide valley in the spectral density, located between the
peak at zero and the bulk of the spectrum, where the density starts to rise
rapidly.

\begin{figure}
\begin{center}
\includegraphics[width=0.8\textwidth,keepaspectratio]{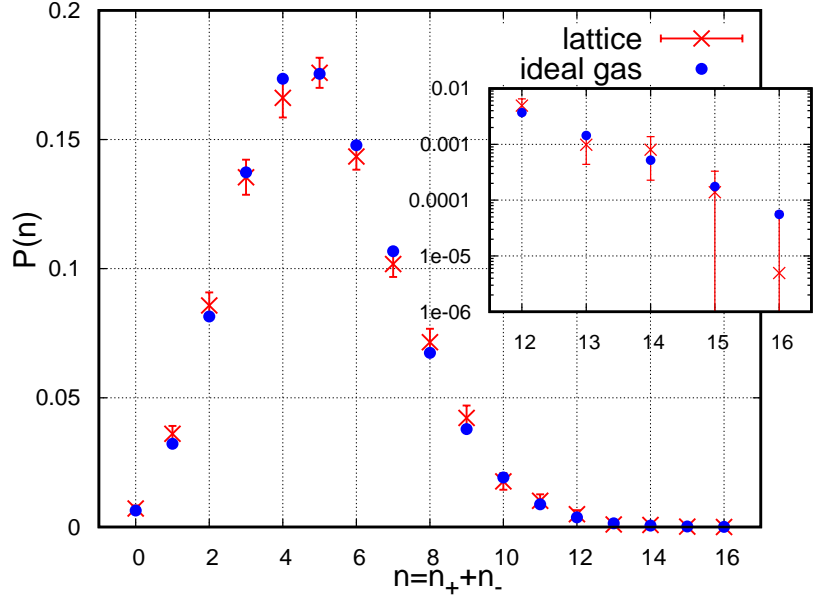}
\caption{\label{fig:ntdist_both} The distribution of the number of eigenmodes
  in the zero mode zone. The red crosses are the lattice data on an ensemble
  of $N_t=8, V=32^3$ lattices at a temperature of $T=1.045T_c$. The blue
  circles represent the expected ideal gas distribution of the number of
  topological objects with their density equal to the density of eigenvalues
  in the zero mode zone. }
\end{center}
\end{figure}

If all these eigenmodes in the zero mode zone are associated to topological
lumps of unit charge that form an ideal gas, then the distribution of the
number of these eigenmodes should follow a Poisson distribution with mean
$\chi V$. We already fixed the mean to $\chi V$, but it is a nontrivial test
whether the distribution is actually Poissonian.  This can be easily checked
in the lattice overlap spectra by counting the number of eigenvalues in the
zero mode zone, configuration by configuration. We show the result in
Fig.~\ref{fig:ntdist_both}. Also shown in the figure is the expected Poisson
distribution, that is the distribution of the number of topological lumps in
an ideal instanton gas with the given susceptibility. Notice that there is no
further fitting involved here, as the only parameter of the distribution,
$\chi V$ has already been determined from the exact zero modes of the
overlap. The agreement with the lattice data is fairly good, indicating that
the topological objects form a non-interacting gas. 

\begin{figure}
\begin{center}
\includegraphics[width=0.8\textwidth,keepaspectratio]{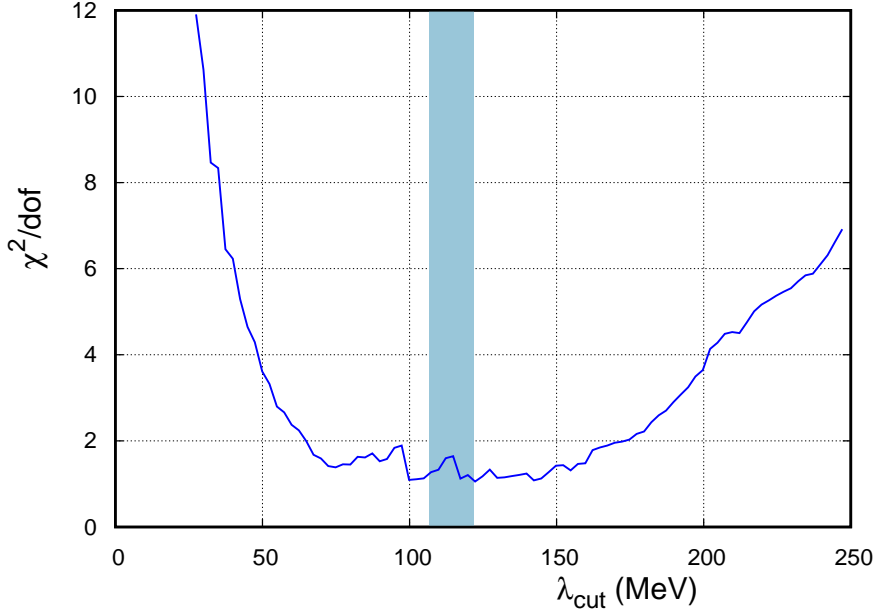}
\caption{\label{fig:cut_vs_chi} The chi squared per degree of freedom of a fit
  to a Poisson distribution of the number of eigenvalues in an interval
  $[-\lambda_{cut},\lambda_{cut}]$ as a function of $\lambda_{cut}$. The blue
  bar represents the previously determined value of $\lambda_{zmz}$ with its
  width being the uncertainty.}
\end{center}
\end{figure}

At this point one could also ask the question whether $\lambda_{zmz}$ is
really a special point of the spectrum in the sense that the number of
eigenvalues in the $[-\lambda_{zmz},\lambda_{zmz}]$ interval is Poisson
distributed, but for other intervals it cannot be described with a Poisson
distribution. To see this, we tried to fit the distribution of the number of
eigenvalues in intervals $[-\lambda_{cut},\lambda_{cut}]$ for different values
of $\lambda_{cut}$. The chi squared per degree of freedom as a function of
$\lambda_{cut}$ is shown in Fig.~\ref{fig:cut_vs_chi}. Indeed, $\lambda_{zmz}$
is in the middle of the spectral region where the chi squared of the fit is
acceptable, and the distribution is close to a Poisson distribution. Notice
that this region is rather wide because of the simple reason that the spectral
density is very small there, and by increasing $\lambda_{cut}$, so few
eigenvalues are added to the data set that they do not influence the quality
of the fit significantly.

\section{Conclusions}

In the present paper we studied the peak at zero virtuality in the spectral
density of the QCD Dirac operator. We saw that the spectral density at the
peak has unexpectedly large finite size corrections. However, this volume
dependence of the spectral density can be understood if we assume that the
eigenmodes in the spectral peak are mixing instanton-antiinstanton zero modes,
originating from a non-interacting gas of topological objects. The
distribution of the number of eigenmodes in the peak also supports this
picture.

The most important consequence of our findings is that there is a strong
connection between the topological susceptibility and the total number of
eigenvalues in the zero mode zone, most of which are in the spectral peak. In
particular, a nonzero topological susceptibility implies the presence of a
corresponding spectral peak, which means that chiral symmetry is not
immediately restored in the high temperature phase.

We have to add some qualifications to the above statement. The numerical
results we presented, were obtained in the quenched approximation. Dynamical
fermions are expected to suppress fluctuations of the topological charge, and
will definitely affect both the topological susceptibility and the spectral
peak. However, as long as the fermions do not induce strong interactions among
topological objects, the connection between the susceptibility and the
spectral peak remains valid. Moreover, even if light dynamical quarks induce
strong interactions among instantons, we expect this effect to suppress the
net topological charge, i.e.\ the susceptibility, more than the near zero modes
in the peak. This means that even in the presence of fermion-induced
interactions, we expect a nonzero topological susceptibility to imply the
presence of a spectral peak. It is a dynamical question how at a given value
of the quark mass the breaking of chiral symmetry due to the spectral peak
compares to the explicit breaking by the quark mass. Presently there is
ongoing work to explore how dynamical quarks influence the spectral peak.

\end{document}